\documentclass[prd,twocolumn,showpacs,preprintnumbers,amsmath,amssymb,superscriptaddress,nofootinbib,english]{revtex4-1}

\usepackage{graphicx}% Include figure files
\usepackage{dcolumn}% Align table columns on decimal point
\usepackage{bm}% bold math
\usepackage{epsfig}
\usepackage{graphicx}
\usepackage{hyperref}
\usepackage[usenames]{color}
\usepackage{url}
\usepackage{enumitem}
\usepackage{float}
\usepackage[utf8]{inputenc}

\hypersetup{
    colorlinks=true,
    linkcolor=blue,
    citecolor=blue,
}

\newcommand{\remove}[1]{}

\def\ie{{\frenchspacing\it i.e.}}
\def\eg{{\frenchspacing\it e.g.}}

\def\be{\begin{equation}}
\def\ee{\end{equation}}
\def\ba{\begin{eqnarray}}
\def\ea{\end{eqnarray}}

\def\rd{r_{\rm d}}
\def\rdh{r_{\rm d}h}
\def\om{\Omega_{\rm m}h^2}
\def\Om{\Omega_{\rm m}}

\begin{document}

\title{Why reducing the cosmic sound horizon alone can not fully resolve the Hubble tension}

\author{Karsten Jedamzik} 
\affiliation{Laboratoire de Univers et Particules de Montpellier, UMR5299-CNRS, Universite de Montpellier, 34095 Montpellier, France}
\email[]{karsten.jedamzik@umontpellier.fr}

\author{Levon Pogosian}
\affiliation{Department of Physics, Simon Fraser University, Burnaby, British Columbia, Canada V5A 1S6}
\email[]{levon@sfu.ca}

\author{Gong-Bo Zhao}
\affiliation{National Astronomy Observatories, Chinese Academy of Science, Beijing, 100101, P.R.China}
\affiliation{University of Chinese Academy of Sciences, Beijing, 100049, P.R.China}
\email[]{gbzhao@nao.cas.cn}

\begin{abstract}
The mismatch between the locally measured expansion rate of the universe and the one inferred from the cosmic microwave background measurements by Planck in the context of the standard $\Lambda$CDM, known as the Hubble tension, has become one of the most pressing problems in cosmology. A large number of amendments to the $\Lambda$CDM model have been proposed in order to solve this tension. Many of them introduce new physics, such as early dark energy, modifications of the standard model neutrino sector, extra radiation, primordial magnetic fields or varying fundamental constants, with the aim of reducing the sound horizon at recombination $r_{\star}$. We demonstrate here that any model which only reduces $r_{\star}$ can never fully resolve the Hubble tension while remaining consistent with other cosmological datasets. We show explicitly that models which achieve a higher Hubble constant with lower values of matter density $\Omega_m h^2$ run into tension with the observations of baryon acoustic oscillations, while models with larger $\Omega_mh^2$ develop tension with galaxy weak lensing data.
\end{abstract}

%\pacs{98.80. -k, 98.80.Es}

\maketitle

Decades of progress in observational and theoretical cosmology have led to the consensus that our universe is well described by a flat Friedman-Robertson-Lemaitre metric and is currently comprised of around $5\%$ baryons, $25\%$ cold dark matter (CDM), and $70\%$ dark energy in its simplest form -- the cosmological constant $\Lambda$. Although this $\Lambda$CDM model fits many observations exquisitely well, its prediction for the present day cosmic expansion rate, $H_0=67.36 \pm 0.54$ km/s/Mpc \cite{Aghanim:2018eyx}, based on precise cosmic microwave background (CMB) radiation observations by the Planck satellite, do not compare well with direct measurements of the Hubble constant. In particular, the Supernovae H0 for the Equation of State (SH0ES) collaboration \cite{Reid:2019tiq}, using Cepheid calibrated supernovae Type Ia, finds a much higher value of $H_0=73.5 \pm 1.4$ km/s/Mpc. This $4.2\sigma$ disagreement, known as the ``Hubble tension'', has spurred much interest in modifications of the $\Lambda$CDM model capable of resolving it (cf. \cite{DiValentino:2020zio} for a comprehensive list of references). Several other determinations of $H_0$, using different methods, are also in some degree of tension with Planck, such as the Megamaser Cosmology Project \cite{Pesce:2020xfe} finding $73.9\pm 3.0$ km/s/Mpc or H0LiCOW \cite{Wong:2019kwg} finding $73.3^{+1.7}_{-1.8}$ km/s/Mpc. It is worth noting that a somewhat lower value of $69.8\pm 2.5$ km/s/Mpc was obtained using an alternative method for calibrating SNIa \cite{Freedman:2019jwv}. 

Among the most precisely measured quantities in cosmology are the locations of the acoustic peaks in the CMB temperature and polarization anisotropy spectra. They determine the angular size of the sound horizon at recombination, 
\be\label{eq:theta} \theta_\star\equiv\frac{r_{\star}}{D(z_{\star})},\ee
with an accuracy of $0.03\%$ \cite{Aghanim:2018eyx}. The sound horizon $r_{\star}$ is the comoving distance a sound wave could travel from the beginning of the universe to recombination, a standard ruler in any given model, and $D(z_{\star})$ is the comoving distance from a present day observer to the last scattering surface, \ie, to the epoch of recombination. $D(z_{\star})$ is determined by the redshift-dependent expansion rate $H(z)=h(z)\times 100$ km/s/Mpc which, in the flat $\Lambda$CDM model, depends only on two parameters (see Appendix~\ref{sec:CMB_BAO} for details): $\Omega_m h^2$ and $h$, where $\Omega_m$ is the fractional matter energy density today and $h=h(0)=H_0/100$ km/s/Mpc. Thus, given $r_{\star}$ and an estimate of $\Omega_m h^2$, one can infer $h$ from the measurement of $\theta_\star$. Using the Planck best fit values of $\Omega_m h^2 = 0.143\pm 0.001$ and $r_\star=144.44 \pm 0.27$ Mpc, obtained within the $\Lambda$CDM model \cite{Aghanim:2018eyx}, yields a Hubble constant significantly lower than the more direct local measurements.

If the value of the Hubble constant was the one measured locally, \ie, $h\approx 0.735$, it would yield a much larger value of $\theta_\star$ unless something else in Eq.~(\ref{eq:theta}) was modified to preserve the observed CMB acoustic peak positions. There are two broad classes of models attempting to resolve this tension by introducing new physics. One introduces modifications at late times (\ie, lower redshifts), \eg, by introducing a dynamical dark energy or new interactions among the dark components that alter the Hubble expansion to make it approach a higher value today, while still preserving the integrated distance $D$ in Eq.~(\ref{eq:theta}). In the second class of models, the new physics aims to reduce the numerator in Eq.~(\ref{eq:theta}), \ie, modify the sound horizon at recombination.

Late time modifications based on simple phenomenological parameterizations tend to fall short of fully resolving the tension \cite{Benevento:2020fev}. This is largely because the baryon acoustic oscillation (BAO) and supernovae (SN) data, probing the expansion in the $0 \lesssim z \lesssim 1$ range, are generally consistent with a constant dark energy density. One can accommodate a higher value of $H_0$ by making parameterizations more flexible, as \eg, in \cite{Zhao:2017cud,Wang:2018fng}, that allow for a non-monotonically evolving effective dark energy fluid. Such non-monotonicity tends to imply instabilities within the context of simple dark energy and modified gravity theories \cite{Zucca:2019ohv} but can, in principle, be accommodated within the general Horndeski class of scalar-tensor theories \cite{Raveri:2019mxg}.

Early-time solutions aim to reduce $r_{\star}$ with essentially two possibilities: (i) a coincidental increase of the Hubble expansion around recombination or (ii) new physics that alters the rate of recombination. Proposals in class (i) include the presence of early dark energy~\cite{Karwal:2016vyq,Poulin:2018cxd,Agrawal:2019lmo,Lin:2019qug, Berghaus:2019cls,Niedermann:2020dwg}, extra radiation in either neutrinos~\cite{Kreisch:2019yzn,Sakstein:2019fmf,Archidiacono:2020yey,Escudero:2019gvw} or some other dark sector~\cite{Anchordoqui:2019yzc,Gonzalez:2020fdy,Pandey:2019plg,Lesgourgues:2015wza,Buen-Abad:2017gxg,Kumar:2018yhh}, and dark energy-dark matter interactions \cite{Agrawal:2019dlm}. Proposals in class (ii) include primordial  magnetic fields~\cite{Jedamzik:2020krr}, non-standard recombination \cite{Chiang:2018xpn}, or varying fundamental constants~\cite{Hart:2019dxi,Sekiguchi:2020teg}. In this work we show that any early-time solution which only changes $r_{\star}$ can never fully resolve the Hubble tension without being in significant tension with either the weak lensing (WL) surveys \cite{Abbott:2017wau,Asgari:2020wuj} or BAO \cite{Alam:2016hwk} observations.

The acoustic peaks, prominently seen in the CMB anisotropy spectra, are also seen as BAO peaks in the galaxy power spectra and carry the imprint of a slightly different, albeit intimately related, standard ruler -- the sound horizon at the ``cosmic drag'' epoch (or the epoch of baryon decoupling), $\rd$, when the photon drag on baryons becomes unimportant. As the latter takes place at a slightly lower redshift than recombination, we have $r_{\rm d} \approx 1.02 r_{\star}$ with the proportionality factor being essentially the same in all proposed modified recombination scenarios. More importantly for our discussion, the BAO feature corresponds to the angular size of the standard ruler at $z \ll z_\star$, \ie, in the range $0 \lesssim z \lesssim 2.5$ accessible by galaxy redshift surveys. For the BAO feature measured using galaxy correlations in the transverse direction to the line of sight, the observable is
\begin{equation}
\theta_\perp^{\rm BAO} (z_{\rm obs}) \equiv \frac{r_{\rm d}}{D(z_{\rm obs})} \, ,
\label{eq:thetaBAO}
\end{equation}
where $z_{\rm obs}$ is the redshift at which a given BAO measurement is made. For simplicity, we do not discuss the line of sight and the ``isotropic'' BAO measurements \cite{Eisenstein:2005su} here, but our arguments apply to them as well. It is well known that BAO measurements at multiple redshifts provide a constraint on $\rdh$ and $\Om$.

In any particular model, $r_{\star}$ (and $\rd$) is a derived quantity that depends on $\om$, the baryon density and other parameters. However, in this work, for the purpose of illustrating trends that are common to all models, {\it we treat $r_\star$ as an independent parameter} and assume that no new physics affects the evolution of the universe after recombination.

\begin{figure}[htbp]  
\includegraphics[scale=0.45]{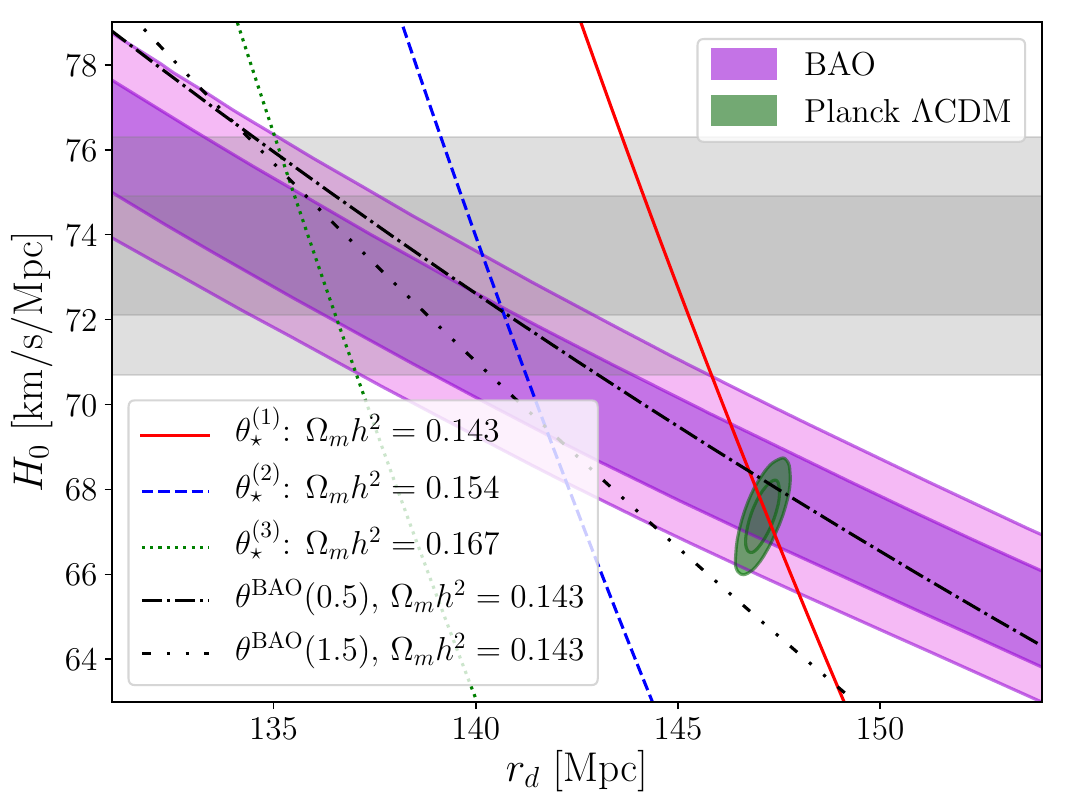}
\caption{ A plot illustrating that achieving a full agreement between CMB, BAO and SH0ES through a reduction of $\rd$ requires a higher value of $\om$. Shown are the lines of degeneracy between the sound horizon $\rd$ and the Hubble constant $H_0$ defined by the CMB acoustic scale $\theta_\star$ at three different values of $\om$: $0.143$, $0.155$ and $0.167$. Also shown are the marginalized $68\%$ and $95\%$ CL bands derived from the combination of all current BAO data, and the $\Lambda$CDM based bounds from Planck. To demonstrate how the slope of the lines changes with redshift, we show two lines corresponding to the SDSS measurements of $\theta_\perp^{\rm BAO}$ at $z=0.51$ and $z=1.5$ \cite{duMasdesBourboux:2020pck} at a fixed $\om=0.143$. The grey band shows the $68\%$ and $95\%$ CL determination of the Hubble constant by SH0ES.
} 
\label{fig:bao-cmb}.
\end{figure}

Without going into specific models, we now consider modifications of $\Lambda$CDM which decrease $r_{\star}$, treating the latter as a free parameter. The relation between $r_\star$ and $\rd$ in different models that reduce the sound horizon is largely the same as the one in $\Lambda$CDM, hence we fix it at $\rd = 1.0184 r_\star$ based on the Planck best fit $\Lambda$CDM value. For a given $\om$, Eq.~(\ref{eq:theta}) defines a line in the $\rd$-$H_0$ plane and, since Eqs. (\ref{eq:theta}) and (\ref{eq:thetaBAO}) are the same in essence, a BAO measurement at each different redshift also defines a respective line in the $\rd$-$H_0$ plane. However, the significant difference between $z_{\star}$ and $z_{\rm obs}$ results in different slopes of the respective $r_{\rm d}(h)$ lines (see Appendix~\ref{sec:fig1} for details), as illustrated in Fig.~\ref{fig:bao-cmb}. The latter shows the $r_{\rm d}(h)$ lines from two different BAO observations, one at redshift $z = 0.5$ and another at $z=1.5$, at $\om$ fixed to the Planck best fit $\Lambda$CDM value of $0.143$, and the analogous lines defined by the CMB acoustic scale plotted for three values of $\om$: $0.143$, $0.155$ and $0.167$. Both lines correspond to transverse BAO measurements. Slopes derived from the line of sight and isotropic BAO at the same redshift would be different, but the trend with increasing redshift is the same. The lines are derived from the central observational values and do not account for the uncertainties in $\theta_\perp^{\rm BAO}$ and $\theta_\star$ (although the uncertainty in $\theta_\star$ is so tiny that it would be difficult to see by eye on this plot). As anticipated, the slope of the $\rd(h)$ lines becomes steeper with increased redshift. 

Also shown in Fig.~\ref{fig:bao-cmb} are the marginalized $68\%$ and $95\%$ confidence levels (CL) derived from the combination of all presently available BAO observations in a recombination-model-independent way, namely, while treating $\rd$ as an independent parameter (see \cite{Pogosian:2020ded} and Appendix~\ref{sec:fig1} for details). The red contours show the $\Lambda$CDM based constraint from Planck, in good agreement with BAO at $H_0\approx 67\,$km/s/Mpc, but in tension with the SH0ES value shown with the grey band. In order to reconcile Planck with SH0ES solely by reducing $\rd$, one would have to move along one of the CMB lines. Doing it along the line at $\om=0.143$ would quickly move the values of $\rd$ and $H_0$ out of the purple band, creating a tension with BAO. Full consistency between the observed CMB peaks, BAO and the SH0ES Hubble constant could only be achieved at a higher value of $\om \approx 0.167$\footnote{The dependence of the CMB $\rd(h)$ lines on $\om$ may appear contradictory to the $\om$ dependence shown in Fig.~1 of a well-know paper by Knox and Millea \cite{Knox:2019rjx}. There, increasing $\om$ moves the CMB best fit ($\rd,h$) point in a direction orthogonal to where our CMB lines move. The reason for the difference is that their $\rd$ is a derived parameter obtained from the standard recombination model and, hence, depends on $\om$. In our derivation of the CMB lines, on the other hand, the $\om$ dependence only appears in $D(z_\star)$ and $D(z_{\rm obs})$.}. However, unless one supplements the reduction in $\rd$ by yet another modification of the model, such high values of $\om$ would cause tension with galaxy WL surveys such as the Dark Energy Survey (DES) \cite{Abbott:2017wau} and the Kilo-Degree Survey (KiDS) \cite{Asgari:2020wuj}, which we illustrate next.

DES and KiDS derived strong constraints on the quantity $S_8 \equiv \sigma_8(\Omega_m/0.3)^{0.5}$, where $\sigma_8$ is the matter clustering amplitude on the scale of $8 \ h^{-1}$Mpc, as well as $\Om$. The value of $S_8$ depends on the amplitude and the spectral index of the spectrum of primordial fluctuations, which are well-determined by CMB and have similar best fit values in all modified recombination models. $S_8$ also depends on the net growth of matter perturbations which increases with more matter, \ie, a larger $\om$. 

\begin{figure}[htbp] 
\includegraphics[scale=0.45]{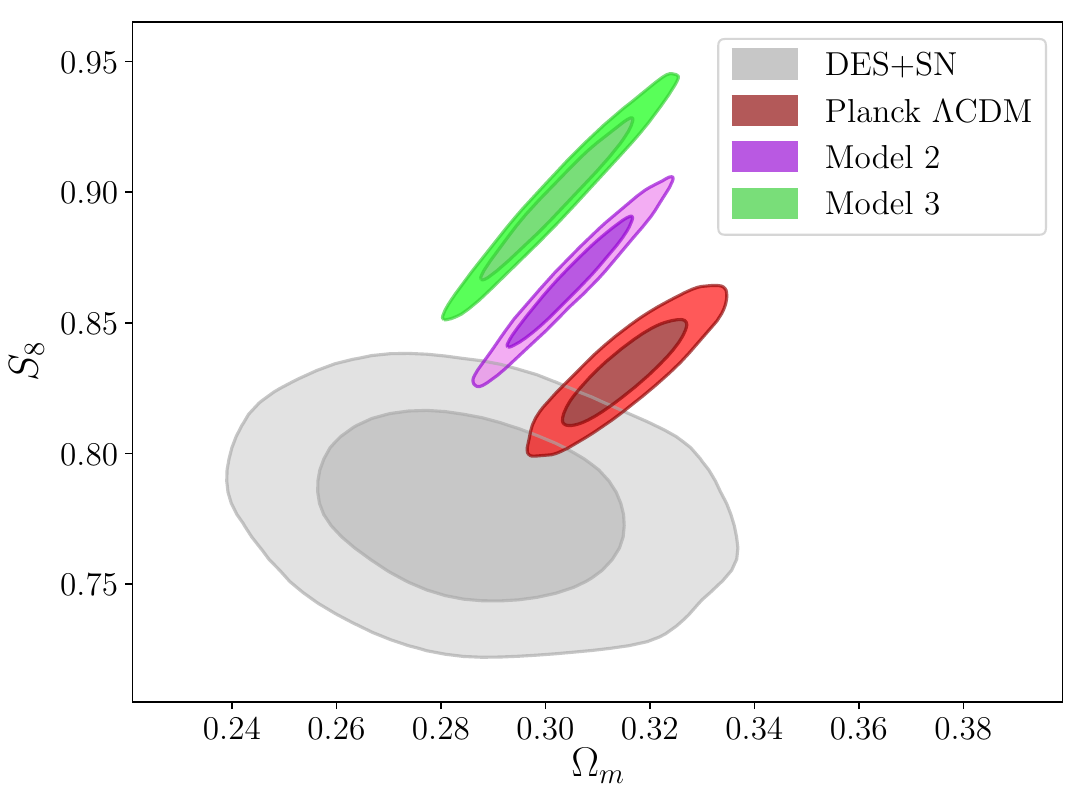}
\caption{
The $68\%$ and $95\%$ confidence level bounds on $S_8$ and $\Om$. Shown are the constraints derived by fitting the $\Lambda$CDM model to a joint dataset of DES and SN and to Planck, along with the contours for Model 2 and Model 3. Model 2 is defined by the simultaneous fit to BAO and CMB acoustic peaks at $\om = 0.155$, {\it i.e.} the overlap between the BAO band and the $\theta_\star^{(2)}$ line in Fig.~\ref{fig:bao-cmb}. Model 3 has $\om=0.167$ and corresponds to the overlap region between the $\theta_\star^{(3)}$ line and the BAO and SH0ES bands in Fig~\ref{fig:bao-cmb}.
} 
\label{fig:S8-Om}.
\end{figure}

The values of $S_8$ and $\Om$ obtained by DES and KiDS are already in slight tension with the Planck best fit $\Lambda$CDM model, and the tension between KiDS and Planck is notably stronger than that between DES and Planck. Increasing the matter density aggravates this tension -- a trend that can be seen in Fig.~\ref{fig:S8-Om}. The figure shows the  $68\%$ and $95\%$ CL joint constraints on $S_8$-$\Om$ by DES supplemented by the Pantheon SN sample \cite{Scolnic:2017caz} (which helps by providing an independent constraint on $\Om$), along with those by Planck within the $\Lambda$CDM model. The purple contours (Model 2) correspond to the model that can simultaneously fit BAO and CMB acoustic peaks at $\om = 0.155$, \ie, the model defined by the overlap between the BAO band and the $\theta_\star^{(2)}$ (blue dashed) line in Fig.~\ref{fig:bao-cmb}. The green contours (Model 3) are derived from the model with $\om=0.167$ corresponding to the overlap region between the $\theta_\star^{(3)}$ (green dotted) line and the BAO and SH0ES bands in Fig.~\ref{fig:bao-cmb} (see Appendix~\ref{sec:fig2} for details). The figure shows that when attempting to find a full resolution of the Hubble tension, with CMB, BAO and SH0ES in agreement with each other, one exacerbates the tension with DES and KiDS.

We note that there is much more information in the CMB than just the positions of the acoustic peaks. It is generally not trivial to introduce new physics that reduces $r_{\star}$ and $\rd$ without also worsening the fit to other features of the temperature and polarization spectra \cite{Knox:2019rjx}. Our argument is that, even if one managed to solve the Hubble tension by reducing $r_\star$ while maintaining a perfect fit to all CMB data, one would still necessarily run into problems with either the BAO or WL.

\begin{figure*}[htbp]  
\includegraphics[scale=0.5]{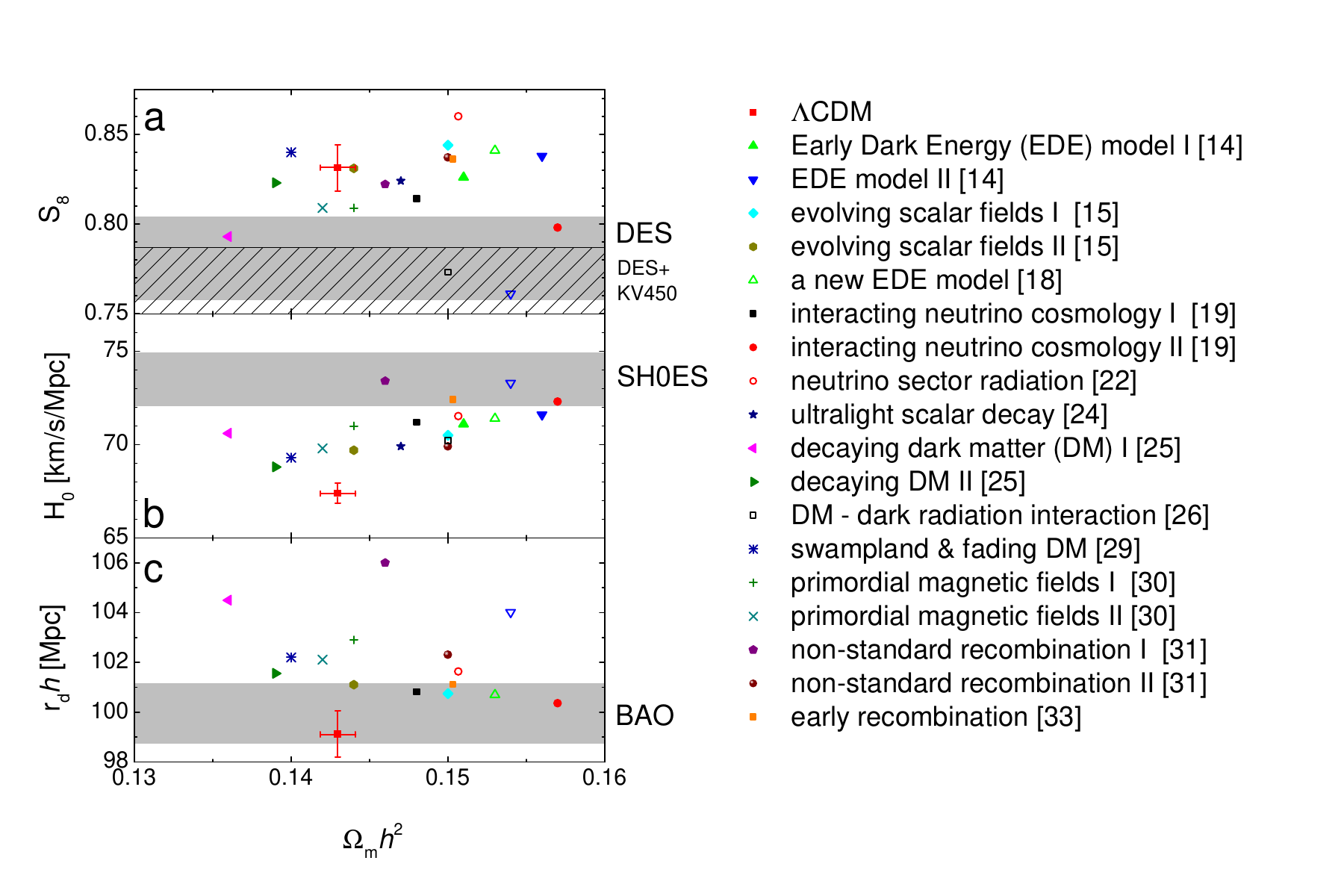}
\caption{
A compilation of values of $\Omega_m h^2$, $\rdh$, $H_0$ and $S_8$ predicted by some of the models aiming to relieve the Hubble tension by lowering the sound horizon. The best fit values of $S_8,H_0,\rdh$ (panels a-c respectively), along with $\Omega_m h^2$ (the horizontal axis), obtained within the models listed on the right. The horizontal bands show the 68\% confidence level observational constraint on the corresponding parameter from different (types of) surveys. The sub-labels a and b in the list of models denote either different choices of model parameters within the same model, or constraints derived from different data combinations on the same model. The red square point with error bars represents the Planck best fit $\Lambda$CDM model \cite{Aghanim:2018eyx}. With the exception of the red dot, corresponding to the model from \cite{Kreisch:2019yzn} with multiple modifications of $\Lambda$CDM fit to Planck temperature anisotropy data only, there is a consistent trend: models with low $\om$ either fail to achieve a sufficiently high $H_0$ or are in tension with baryonic acoustic oscillations (BAO), and models with high values of $\om$ run into tension with DES or KiDS.
} 
\label{fig:models}.
\end{figure*}

Surveying the abundant literature of the proposed early-time solutions to the Hubble tension, one finds that the above trends are always confirmed. Fig.~\ref{fig:models} shows the best fit values of $\rdh$, $H_0$ and $S_8$ in models from Refs. \cite{Kreisch:2019yzn,Poulin:2018cxd,Pandey:2019plg,Agrawal:2019lmo,Gonzalez:2020fdy,Chiang:2018xpn,Jedamzik:2020krr,Agrawal:2019dlm,Sekiguchi:2020teg}. Note that there are other proposed early-time solutions to the Hubble tension. Fig.~\ref{fig:models} only shows the models for which explicit estimates of $H_0$, $\Omega_mh^2$, $S_8$, and possibly $\rdh$ were provided. One can see that, except for the model represented by the red dot at the very right of the plot, corresponding to the strongly interacting neutrino model of \cite{Kreisch:2019yzn}, solutions requiring low $\om$ are in tension with BAO, whereas solutions with higher $\om$ are in tension with DES and KiDS. This latter tension was previously observed and extensively discussed in the context of the early dark energy models \cite{Hill:2020osr,Ivanov:2020ril,DAmico:2020ods,Niedermann:2020qbw,Murgia:2020ryi,Smith:2020rxx}. As we have shown in this paper, it is part of a broader problem faced by all proposals aimed at reducing the Hubble tension in which the main change amounts to a reduction of $\rd$.

In most of the models represented in Fig.~\ref{fig:models}, the effect of introducing new physics only amounts to a reduction in $\rd$. We note that, in any specific model of a reduced $\rd$, the best fit values of other cosmological parameters also change, which can affect the quality of the fit to various datasets. However, such changes, {\it e.g.} in the best fit value of the spectral index $n_s$ which affects $S_8$, tend to be small for the models studied in the literature and have a minor impact compared to the effect of reducing $\rd$, which is a pre-requisite for reconciling CMB with SH0ES. As we have argued, this will necessarily limit their ability to address the Hubble tension while staying consistent with the large scale structure data. Resolving the Hubble tension by new early-time physics without creating other observational tensions requires more than just a reduction of the sound horizon. This is exemplified by the interacting dark matter-dark radiation model \cite{Lesgourgues:2015wza} and the neutrino model \cite{Kreisch:2019yzn} proposed as solutions. Here, extra tensions are avoided by supplementing the reduction in the sound horizon due to extra radiation by additional exotic physics: dark matter-dark radiation interactions in the first case and neutrino self-interactions and non-negligible neutrino masses in the second case. Consequently, with so many parameters, the posteriori probabilities for cosmological parameters are highly inflated over those for $\Lambda$CDM. It is not clear how theoretically appealing such scenarios are, and the model in \cite{Kreisch:2019yzn} seems to be disfavoured by the CMB polarization data.

In conclusion, we have argued that any model which tries to reconcile the CMB inferred value of $H_0$ with that measured by SH0ES by only reducing the sound horizon automatically runs into tension with either the BAO or the galaxy WL data. While we do not expect our findings to be surprising for the majority of the community, the novelty of our result is in isolating and clearly stating the essence of the problem -- that the slopes of the $r_\star$-$H_0$ degeneracy lines for BAO and CMB are vastly different, thus making it impossible to reconcile CMB with SH0ES by reducing $r_\star$ without violating BAO. We believe this very simple fact has not been stated before in this context in a model-independent way. With just a reduction of $r_{\star}$, the highest value of the Hubble constants one can get, while remaining in a reasonable agreement with BAO and DES/KiDS, is around $70$ km/s/Mpc. Thus, a full resolution of the Hubble tension will require either multiple modifications of the $\Lambda$CDM model or discovering systematic effects in one or more of the datasets.

\acknowledgments

We thank Eiichiro Komatsu and Joulien Lesgourgues for helpful comments on the draft of the paper and Kanhaiya Pandy and Toyokazu Sekiguchi for kindly providing us with data of their models. We gratefully acknowledge using {\tt CosmoMC} \cite{Lewis:2002ah} and {\tt GetDist} \cite{Lewis:2019xzd}. This research was enabled in part by support provided by WestGrid ({\tt www.westgrid.ca}) and Compute Canada Calcul Canada ({\tt www.computecanada.ca}). L.P. is supported in part by the National Sciences and Engineering Research Council (NSERC) of Canada, and by the Chinese Academy of Sciences President's International Fellowship Initiative, Grant No. 2020VMA0020. G.B.Z. is supported by the National Key Basic Research and Development Program of China (No. 2018YFA0404503), a grant of CAS Interdisciplinary Innovation Team, and NSFC Grants 11925303, 11720101004, 11673025 and 11890691.
 
\appendix

\section{The acoustic scale measurements from the CMB and BAO}
\label{sec:CMB_BAO}

The CMB temperature and polarization anisotropy spectra provide a very accurate measurement of the angular size of the sound horizon at recombination,
\begin{equation}
\theta_\star = \frac{r_{\star}}{D(z_{\star})} 
\, ,
\label{eq:theta*}
\end{equation}
where $r_\star$ is the sound horizon at recombination, or the comoving distance a sound wave could travel from the beginning of the universe to recombination, and $D(z_{\star})$ is the comoving distance from a present day observer to the last scattering surface, \ie, to the epoch of recombination. In a given model, $r_\star$ and $D(z_{\star})$ can be determined from $r_{\star} = \int_{z_{\star}}^\infty c_s(z) {\rm d} z /H(z)$ and $D(z_{\star})=\int^{z_\star}_0 c \ {\rm d} z / H(z)$, where $c_s(z)$ is the sound speed of the photon-baryon fluid, $H(z)$ is the redshift-dependent cosmological expansion rate and $c$ is the speed of light. To complete the prescription, one also needs to determine $z_\star$ using a model of recombination.

The redshift dependence of the Hubble parameter in the $\Lambda$CDM model can be written as
\begin{equation}
h(z) = \sqrt{\Omega_r h^2 (1+z)^4 + \Omega_m h^2 (1+z)^3 + \Omega_{\Lambda} h^2}
\label{eq:hz_lcdm}
\end{equation}
where $h(z)$ is simply $H(z)$ in units of 100 km/s/Mpc, and $h$ is the value at redshift $z=0$. Here, $\Omega_r$, $\Omega_m$ and $\Omega_{\Lambda}$ are the present day density fractions of radiation, matter (baryons and CDM) and dark energy. From the precise measurement of the present-day CMB temperature $T_0 = 2.7255\,$K (however, also see \cite{Ivanov:2020mfr}), and adopting the standard models of particle physics and cosmology, one knows the density of photons and neutrinos $\Omega_r h^2$. Using the theoretically well motivated criticality condition on the sum of the fractional densities, \ie, $\Omega_r + \Omega_m + \Omega_{\Lambda} = 1$, one finds that $h(z)$ is dependent only on two remaining quantities: $\Omega_m h^2$ and $h$. The photon-baryon sound speed $c_s$ in Eq.~(\ref{eq:theta}) is determined by the ratio of the baryon and photon densities and is well-constrained by both Big Bang nucleosynthesis and the CMB. Fitting the $\Lambda$CDM model to CMB spectra also provides a tight constraint on $\om$, making it possible to measure $h$.

In alternative models, a smaller $r_\star$ is achieved by introducing new physics that reduces $z_\star$ through a modification of the recombination process or by modifying $h(z)$ before and/or during recombination, or a combination of the two. In our analysis, we consider Eq.~(\ref{eq:theta*}) while remaining agnostic about the particular model that determines the sound horizon. Namely, {\it we treat $r_\star$ as an independent parameter}. We assume, however, that after the recombination, the expansion of the universe is well-described by Eq.~(\ref{eq:hz_lcdm}), which is the case in many alternative models. Thus, our independent parameters are $r_\star$, $\om$ and $h$, with the latter two determining $D(z_{\star})$. The dependence of $D(z_{\star})$ on the precise value of $z_\star$ is very weak, so that the differences in $z_\star$ in different models do not play a role.

The same acoustic scale is also imprinted in the distribution of baryons. There are three types of BAO observables corresponding to the three ways of extracting the acoustic scale from galaxy surveys \cite{Eisenstein:2005su}: using correlations in the direction perpendicular to the line of sight, using correlations in the direction parallel to the line of sight, and the angle-averaged or ``isotropic'' measurement. While our MCMC analysis includes all three types of the BAO data, for the purpose of our discussion it suffices to consider just the first type, which is the closest to CMB in its essence, but our conclusions apply to all three. Namely, we consider 
\begin{equation}
\theta_\perp^{\rm BAO} (z_{\rm obs}) \equiv \frac{r_{\rm d}}{D(z_{\rm obs})} 
\, ,
\label{eq:thetaBAO*}
\end{equation}
where $\rd=\int_{z_{\rm d}}^\infty c_s(z){\rm d} z/H(z)$ is the sound horizon at the epoch of baryon decoupling, closely related to $r_\star$, and $z_{\rm obs}$ is the redshift at which a given BAO measurement is made. We adopt a fixed relation $\rd = 1.0184 r_\star$ that holds for the Planck best fit $\Lambda$CDM model and is largely unchanged in the alternative models.

As the distance integrals $D(z_\star)$ and $D(z_{\rm obs})$ in the denominators of Eqs. (\ref{eq:theta*}) and (\ref{eq:thetaBAO*}) are dominated by the matter density at low redshifts, one can safely neglect $\Omega_r h^2$ and write
\begin{equation}
\theta_{\star} = {r_\star \over 2998\, {\rm Mpc}} \left( \int_0^{z_{\star}}
\frac{{\rm d}z}{\omega_m^{1/2}\sqrt{(1+z)^3 + h^2/\omega_m -1}} \right)^{-1},
\label{eq:rstar}
\end{equation}
where $\omega_{\rm m} = \om$ and $2998\, {\rm Mpc} = c/100{\rm km/s/Mpc}$, and an analogous equation for BAO with the replacement $(r_{\star}, \theta_\star ,z_{\star}) \to (r_{\rm d},\theta_\perp^{\rm BAO},z_{\rm obs})$. For a given $\om$, Eq.~(\ref{eq:rstar}) defines a line in the $\rd$-$H_0$ plane. Similarly, a BAO measurement at each different redshift also defines a respective line in the $\rd$-$H_0$ plane. Taking the derivative of $r_{\star}$ with respect to $h$ one finds
\begin{equation}
\frac{\partial r_{\star}}{\partial h} = 
-\frac{h}{\omega_m}\theta_\star
\int_0^{z_{\star}}
\frac{2998\, {\rm Mpc}\, {\rm d}z}{\omega_m^{1/2}\bigl({(1+z)^3 + h^2/\omega_m -1}\bigr)^{3/2}}
\label{eq:drdh}
\end{equation}
and a completely analogous equation for BAO. It is important to realize that the derivative is very different for CMB and BAO due to the vast difference in redshifts at which the standard ruler is observed, $z_{\star}\approx 1100$ for CMB {\it vs} $z_{\rm obs} \sim 1$ for BAO, resulting in different values of the integral in Eq. (\ref{eq:drdh}). This results in different slopes of the respective $r_{\rm d}(h)$ lines. Note that the slopes of the $\rd(h)$ lines differ for the transverse, parallel and volume averaged BAO measured at the same redshift. While important for constraining cosmological parameters \cite{Addison:2017fdm}, these differences are small compared to that caused by the big difference between the BAO and CMB redshifts.

\section{Obtaining the contours and the $\rd(h)$ lines in Fig.~\ref{fig:bao-cmb}}
\label{sec:fig1}

The marginalized joint $\rd-H_0$ constraints from BAO were obtained using {\tt CosmoMC} \cite{Lewis:2002ah} modified to work with $\rd$ as an independent parameter. The cosmological parameters we vary are $\rd$, $\om$ and $h$, and the shown constraint is obtained after marginalizing over $\om$. The BAO data included the recently released Date Release (DR) 16 of the extended Baryon Oscillation Spectroscopic Survey (eBOSS) \cite{Alam:2020sor} that includes BAO and redshift space distortions (RSD) measurements at multiple redshifts from the samples of Luminous Red Galaxies (LRGs), Emission Line Galaxies (ELGs), clustering quasars (QSOs), and the Lyman-$\alpha$ forest. We use the BAO measurement from the full-shape auto- and cross-power spectrum of the eBOSS LRGs and ELGs \cite{Zhao:2020tis,Wang:2020tje}, the BAO measurement from the QSO sample \cite{Hou:2020rse}, and from the Lyman-$\alpha$ forest sample \cite{duMasdesBourboux:2020pck}. We combine these with the low-$z$ BAO measurements by 6dF \cite{Beutler:2011hx} and the SDSS DR7 main Galaxy sample (MGS) \cite{Ross:2014qpa}.

The CMB and BAO lines shown in Fig.~\ref{fig:bao-cmb} were obtained by talking the measured value of $\theta_\star$ or $\theta_\perp(z_{\rm obs})$, fixing $\om$ at a certain value (provided for each line in the legend), varying $h$ and deriving $\rd$ from Eqs.~(\ref{eq:theta*}) and (\ref{eq:thetaBAO*}). We do not show the uncertainties around the individual lines because they are only meant to demonstrate the differences in slopes and the effect of different $\om$. The marginalized BAO and the Planck CMB contours provide a more accurate representation of the uncertainties involved.

\section{Obtaining the $S_8$ constraints in Fig.~\ref{fig:S8-Om}}
\label{sec:fig2}
The joint DES+SN contours in Fig.~\ref{fig:S8-Om} are obtained using the default version of {\tt CosmoMC} and marginalizing over all relevant $\Lambda$CDM and nuisance parameters. To derive the Model 2 and Model 3 contours in Fig.~\ref{fig:S8-Om}, we fit the $\Lambda$CDM model to the BAO data using $\rd$, $\om$ and $h$ as a free parameters, supplemented by Gaussian priors on $\om$ and $h$, and with the primordial spectrum amplitude $A_s$ and the spectral index $n_s$ fixed to their best fit $\Lambda$CDM values. The fit then generates constraints on $S_8$ and $\Om$ as derived parameters. For Model 2, the Gaussian priors were $\om=0.155 \pm 0.0012$, where we assumed the same relative uncertainty in $\om$ as for the Planck best fit $\Lambda$CDM model, and $h=0.71 \pm 0.01$, corresponding to the central value and the 1$\sigma$ overlap between the CMB2 line and the BAO band. For Model 3, the priors were $\om = 0.167 \pm 0.0013$ and $h=0.735 \pm 0.14$.

%\bibliography{h0}


\end{document}